\documentclass{emulateapj}
\usepackage{graphicx}

\begin{document}

\def\etal{et al.\ \rm}
\def\ba{\begin{eqnarray}}
\def\ea{\end{eqnarray}}
\def\etal{et al.\ \rm}
\def\Fdw{F_{\rm dw}}
\def\Tex{T_{\rm ex}}
\def\Fdis{F_{\rm dw,dis}}
\def\Fnu{F_\nu}
\def\FJ{F_{J}}
\def\FJE{F_{J,{\rm Edd}}}

\title{Secular dynamics of the triple system harbouring 
PSR J0337+1715 and implications for the origin of its orbital
configuration.}

\author{Roman R. Rafikov\altaffilmark{1}}
\altaffiltext{1}{Department of Astrophysical Sciences, 
Princeton University, Ivy Lane, Princeton, NJ 08540; 
rrr@astro.princeton.edu}


\begin{abstract}
We explore secular dynamics of a recently discovered hierarchical 
triple system consisting of the radio pulsar PSR J0337+1715 
and two white dwarfs (WDs). We show that three body interactions endow 
the inner binary with a large forced eccentricity and  
suppress its apsidal precession, to about $24\%$ of the rate 
due to the general relativity. However,
precession rate is still quite sensitive to the 
non-Newtonian effects and may be used to constrain 
gravity theories if measured accurately. Small 
value of the free eccentricity of the inner binary 
$e_{i}^{\rm free}\approx 2.6\times 10^{-5}$ and vanishing 
forced eccentricity of the outer, relatively eccentric 
binary naturally result in 
their apsidal near-alignment. In addition, this triple system 
provides a unique opportunity to explore 
excitation of both eccentricity and inclination in neutron 
star-WD (NS-WD) binaries, e.g. due to random torques caused 
by convective eddies in the WD progenitor. We show this process to be highly  
anisotropic and more effective at driving eccentricity rather than 
inclination. The outer binary eccentricity as well as 
$e_{i}^{\rm free}$ exceed by more than an order of magnitude the 
predictions of the eccentricity-period relation of Phinney (1992),
which is not uncommon. We also argue that the non-zero mutual 
inclination of the two binaries emerges at the end 
of the Roche lobe overflow of the outer (rather than the inner) 
binary. 
\end{abstract}

\keywords{stellar dynamics --- celestial mechanics --- 
binaries (including multiple): close --- white dwarfs --- 
pulsars: individual (PSR J0337+1715)}


\section{Introduction.}  
\label{sect:intro}

Recent discovery of a triple system harboring two white dwarfs 
(WDs) and a millisecond pulsar PSR J0337+1715 (Ransom \etal 2014) is 
intriguing because of the complicated evolutionary history that
this system must have undergone (Tauris \& van den Heuvel 2014, hereafter
TvdH14). 
It also represents an interesting testbed of the Newtonian 
dynamics which can be explored at very high accuracy. 

In particular, high masses of the WD companions make
gravitational three body effects quite significant. A pulsar 
for which such study has 
been done previously is PSR 1257+12 (Wolszczan \& Frail 1992) but it is 
orbited by three planetary mass objects (Rasio \etal 1992, Malhotra 
1993). 

In addition, high timing accuracy of PSR J0337+1715 (Ransom \etal 2014)
makes this system well suited for studying orbital 
dynamics. Mutual gravitational interactions 
between the orbiting bodies are also being explored in systems 
of multiple extrasolar transiting planets discovered 
by {\it Kepler} satellite, using the so-called {\it transit timing 
variations} (TTVs; e.g. Mazeh \etal 2013). However, planetary transits
usually have timing accuracy of order minutes, while PSR J0337+1715
already yields median arrival time uncertainty of $0.8\mu$s in 10 
s integrations. Moreover, pulsar timing has the benefit of 
revealing the variations of orbital parameters 
over the {\it full orbit}, while TTVs inform us only about the 
system parameters at the moments of mutual conjunctions of 
eclipsing bodies. 

Our present goal is to explore secular evolution of the 
PSR J0337+1715 system and to provide possible connections to
its origin. This system is hierarchical in 
nature, with the semi-major axis of the outer binary
$a_o=1.76498\times 10^{13}$ cm far exceeding that of 
the inner binary $a_i=4.776\times 10^{11}$ cm. 

The two features of its present day dynamical architecture are 
rather intriguing. First, the orbits of the inner and outer 
binaries are highly coplanar: their mutual inclination 
$i=1.2\times 10^{-2\circ}$ is small but is certainly non zero. 
Second, the orbital ellipses of the inner and outer orbits 
are quite well aligned: the difference of their apsidal angles is 
small, $\varpi_i-\varpi_o=1.9987^\circ$. Understanding 
the origin of these dynamical peculiarities of the PSR J0337+1715 
system will be one of the goals of this work.


\section{General setup.}  
\label{sect:setup}

We employ the notation in which all quantities 
related to pulsar, inner and outer binaries (or WDs) have 
subscripts ``p'', ``i'', and ``o'' correspondingly.
Following conventional approach (adopted in Ransom \etal 2014)
we assume the inner binary to consist of the pulsar and the 
short-period WD, while the outer binary refers to the motion 
of the outer WD w.r.t. the barycenter of the inner binary.

We use the parameters of the system presented in
Ransom \etal (2014): pulsar and WD masses, $m_p=1.4378$M$_\odot$,
$m_i=0.1975$M$_\odot$, $m_o=0.41$M$_\odot$, total mass 
$m_3=2.045$M$_\odot$, orbital periods of the 
inner and outer binaries $P_i=1.6294$ d and $P_o=327.26$ d, 
eccentricities $e_i=6.918\times 10^{-4}$ and 
$e_o=3.5356\times 10^{-2}$.


\section{Secular evolution.}  
\label{sect:evolve}

Unlike the PSR 1257+12 and multi-planet systems with largest TTVs 
observed by {\it Kepler}, PSR J0337+1715 is a hierarchical triple 
and is devoid of mean motion resonances, simplifying 
interpretation of its dynamics. For clarity, in this work we 
also ignore the short-period gravitational perturbations and 
focus on understanding the current
configuration of the system in the {\it secular} approximation 
(Murray \& Dermott 1999), which consists of averaging the 
potentials of all orbiting bodies over their relatively short 
orbital periods. In this approximation the semi-major axes 
of the binaries stay constant on timescales longer than $P_i$
and $P_o$. Also, in the limit $e_i,e_o,i\ll 1$ appropriate for 
PSR J0337+1715 system the evolution of the orbital eccentricities
decouples from the inclination evolution.

In this work we focus on the eccentricity evolution. It is well 
known (Murray \& Dermott 1999) that in secular approximation 
the components of the eccentricity vectors ${\bf e}_i=(k_i,h_i)=
(e_i\cos\varpi_i,e_i\sin\varpi_i)$ and ${\bf e}_o=(k_o,h_o)=
(e_o\cos\varpi_o,e_o\sin\varpi_o)$ of the inner and outer binaries
vary in time as
\ba
h_i(t)=e_{i,+}\sin\left(g_+ t+\beta_+\right)+
e_{i,-}\sin\left(g_- t+\beta_-\right),
\label{eq:h_i}\\
k_i(t)=e_{i,+}\cos\left(g_+ t+\beta_+\right)+
e_{i,-}\cos\left(g_- t+\beta_-\right),
\label{eq:k_i}\\
h_o(t)=e_{o,+}\sin\left(g_+ t+\beta_+\right)+
e_{o,-}\sin\left(g_- t+\beta_-\right),
\label{eq:h_o}\\
k_o(t)=e_{o,+}\cos\left(g_+ t+\beta_+\right)+
e_{o,-}\cos\left(g_- t+\beta_-\right).
\label{eq:k_o}
\ea
Here $g_\pm$ and $e_{i,\pm},e_{o,\pm}$ are the eigenvalues and
eigenvectors of the matrix ${\bf A}$ given by 
\ba
{\bf A} =  
\left(
\begin{array}{ll}
A_i & B_i\\
B_o & A_o
\end{array}
\right)
\ea 
so that 
\ba
g_\pm=\frac{1}{2}\left[A_i+A_o\pm
\sqrt{\left(A_i-A_o\right)^2+4B_iB_o}\right],
\label{eq:g_def}
\ea
and 
\ba
\frac{e_{i,\pm}}{e_{o,\pm}}=
-\frac{B_i}{A_i-g_\pm}.
\label{eq:eigen}
\ea

Because the components of the triple have comparable masses we 
use the elements of matrix ${\bf A}$ from Ford \etal (2000):
\ba
A_i & = & A_i^{\rm sec}+\dot\varpi_{\rm GR} 
\label{eq:A_in}\\
& \approx & 
2.3403\times 10^{-10}~\mbox{s}^{-1}=0.423^\circ\mbox{yr}^{-1},
\nonumber\\
A_i^{\rm sec} & = & 
\frac{3}{4}\sqrt{G}\frac{a_i^{3/2}}{a_0^3}\frac{m_o}{\sqrt{m_p+m_i}}
\label{eq:A_in_sec}\\
& \approx &  1.6634\times 10^{-10}~\mbox{s}^{-1}= 0.3^\circ\mbox{yr}^{-1},
\nonumber\\
\dot\varpi_{\rm GR} & = & 3\frac{G^{3/2}\left(m_p+m_i\right)^{3/2}}
{a_i^{5/2}c^2}
\label{eq:omGR}\\
& \approx &  6.769\times 10^{-11}~\mbox{s}^{-1}=0.122^\circ\mbox{yr}^{-1},
\nonumber\\
A_o & = & \frac{3}{4}\sqrt{G}\frac{a_i^2}{a_0^{7/2}}\frac{m_p m_i
m_3^{1/2}}{\left(m_p+m_i\right)^2}
\label{eq:A_out}\\
& \approx &  1.2958\times 10^{-11}~\mbox{s}^{-1}=0.0234^\circ\mbox{yr}^{-1},
\nonumber\\
B_i & = & -\frac{15}{16}\sqrt{G}\frac{a_i^{5/2}}{a_0^4}
\frac{m_o\left(m_p-m_i\right)}{\left(m_p+m_i\right)^{3/2}}
\label{eq:B_in}\\
& \approx &  -4.267488\times 10^{-12}~\mbox{s}^{-1}=
-7.71\times 10^{-3~\circ}\mbox{yr}^{-1},
\nonumber\\
B_o & = & -\frac{15}{16}\sqrt{G}\frac{a_i^3}{a_0^{9/2}}
\frac{m_p m_i\left(m_p-m_i\right)m_3^{1/2}}
{\left(m_p+m_i\right)^3}
\label{eq:B_out}\\
& \approx & -3.3245\times 10^{-13}~\mbox{s}^{-1}=
-6\times 10^{-4~\circ}\mbox{yr}^{-1}.
\nonumber
\ea
The diagonal element $A_i$ of this matrix for the inner WD includes
not only the secular contribution $A_i^{\rm sec}$ due to gravitational 
coupling to the outer WD but also the general relativistic
precession term $\dot\varpi_{\rm GR}$. All other components of 
${\bf A}$ are secular in nature.

With these values we obtain
\ba
g_+\approx 2.34\times 10^{-10}~\mbox{s}^{-1},~~
g_-\approx 1.295\times 10^{-11}~\mbox{s}^{-1},
\label{eq:g_pm}
\ea
and secular periods corresponding to these eigenvalues are 
\ba
P_+\approx 851~\mbox{yr},~~~
g_-\approx 15,382~\mbox{yr}.
\label{eq:P_pm}
\ea
This numerical estimates clearly illustrate that $g_+\approx A_i$ and
$g_-\approx A_o$ because of the smallness of $B_i$ and $B_o$.

Expressions (\ref{eq:h_i})-(\ref{eq:k_o}) depend on six free 
parameters --- $\beta_\pm$, $e_{i,\pm}$, $e_{o,\pm}$ --- but 
two can be eliminated using the relations (\ref{eq:eigen}). 
Remaining four parameters are fixed using the knowledge
of $k_i,h_i,k_o,h_0$ at a particular moment of time. We use
the values provided in Ransom \etal (2014), $k_i=-9.17\times 10^{-5}$, 
$h_i=6.857\times 10^{-4}$, $k_o=-3.46\times 10^{-3}$, 
$h_o=3.52\times 10^{-2}$, which 
we assume to correspond to time $t=0$. As a result we find
\ba
e_{i,+} & \approx & 2.571
\times 10^{-5},~~~
e_{i,-} \approx 0.0006825,
\label{eq:ei}\\
e_{o,+} & \approx & -3.8657
\times 10^{-8},~~~
e_{o,-} \approx 0.035356,
\label{eq:eo}\\
\beta_+ & \approx & 165.42753^\circ,
~~~\beta_- \approx 95.61955133^\circ.
\label{eq:betas}
\ea
These parameters fully specify secular behavior of the system 
via equations (\ref{eq:h_i})-(\ref{eq:k_o}).


\section{Inner binary}  
\label{sect:inner}

\begin{figure}
\plotone{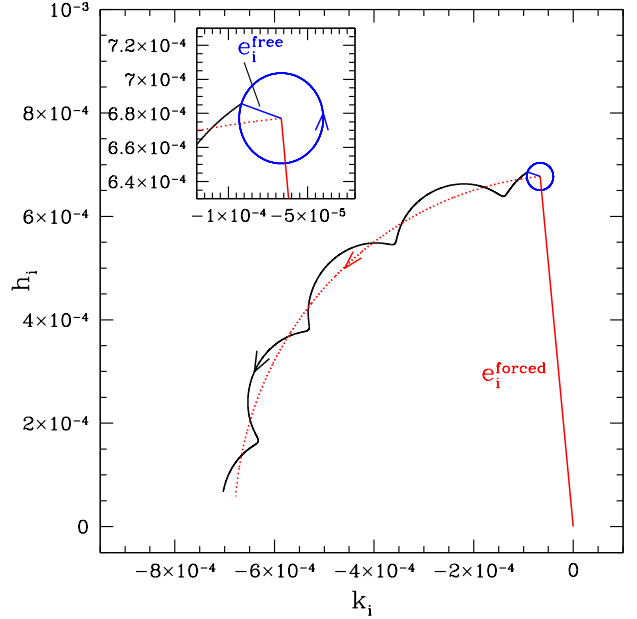}
\caption{Evolution of the inner binary eccentricity vector ${\bf e}_i$ 
(black), decomposed into the forced (red) and free (blue) 
contributions, for $3,400$ yr from the present time. Inset 
focuses on the geometry of the ${\bf e}_i^{\rm free}$ precession,
assuming that its guiding center is not moving. 
The values of ${\bf e}_i^{\rm free}$ and 
${\bf e}_i^{\rm forced}$ at present epoch are shown and 
blue and red line segments. Note that 
$|{\bf e}_i^{\rm free}| \ll |{\bf e}_i^{\rm forced}|$.
\label{fig:inner_prec}}
\end{figure}

Solution for the eccentricity vector of the inner binary
can be represented as the sum of the {\it free} and {\it forced} 
components
\ba
{\bf e}_i & = & {\bf e}_i^{\rm free}+{\bf e}_i^{\rm forced},
\label{eq:in_decomp}\\
{\bf e}_i^{\rm free} & = & e_{i,+}\left\{
\begin{array}{l}
\cos\left(g_+ t+\beta_+\right) \\
\sin\left(g_+ t+\beta_+\right)
\end{array}
\right\},
\label{eq:in_free}\\
{\bf e}_i^{\rm forced} & = & e_{i,-}\left\{
\begin{array}{l}
\cos\left(g_- t+\beta_-\right) \\
\sin\left(g_- t+\beta_-\right)
\end{array}
\right\}.
\label{eq:in_forced}
\ea
The amplitude of the free eccentricity vector is directly 
related to the initial conditions and characterizes the 
``own'' eccentricity of the inner binary. Secular precession
of ${\bf e}_i^{\rm free}$ at the rate $g_+$, as well as the 
forced eccentricity ${\bf e}_i^{\rm forced}$, 
arise because of the gravity of the outer WD. They 
would be absent if the inner binary were isolated. 

This decomposition is illustrated in Figure \ref{fig:inner_prec} where 
we show the forced eccentricity vector at $t=0$ (red) 
and its evolution as $t$ goes up to $3,400$ yr (red dashed) 
in $k-h$ space. Vector 
${\bf e}_i^{\rm free}$ is shown in blue at $t=0$ (see inset) and its 
evolution {\it if ${\bf e}_i^{\rm forced}$ were fixed} is given by the
blue circle. Each of these vectors uniformly circulates
at frequencies $g_+$ and $g_-$ correspondingly. With evolving 
${\bf e}_i^{\rm forced}$, their superposition (black) exhibits a cycloid-like 
epicyclic motion with frequency $g_+$ along the guiding center following 
the big circle made by ${\bf e}_i^{\rm forced}$ with angular frequency 
$g_-$. This motion of ${\bf e}_i$ results in {\it non-uniform} evolution 
of both $k_{i}$ and $h_{i}$, as well as of $e_{i}$ (the distance 
between the origin and a point on the black curve) and of $\varpi_{i}$ 
--- angle between the abscissa axis and ${\bf e}_i$. 

Note that for the inner binary $e_i^{\rm free}\ll e_i^{\rm forced}$, 
since $|e_{i,+}/e_{i,-}|\approx 0.038$. This has 
several important implications discussed further. 

Nonuniform evolution of the orbital elements of the inner WD is 
illustrated in Figure \ref{fig:inner_prec}
where we plot $k_{i}$, $h_{i}$, $e_{i}$, and $\dot \varpi_{i}$ 
as a functions of time. The latter is computed via 
$\dot\varpi=(k\dot h-h\dot k)/(k^2+h^2)$ and varies considerably 
because $g_-/g_+\approx 0.055$ and $|e_{i,+}/e_{i,-}|$ have
comparable magnitudes. 

The present day value of the precession rate for the inner WD is
\ba
\dot\varpi_i\approx 0.02903^\circ~\mbox{yr}^{-1},
\label{eq:dvarpi_i}
\ea
which is less than either $\dot\varpi_{\rm GR}$ or $g_+$. This 
is because of the small value of $e_i^{\rm free}/e_i^{\rm forced}$, 
which suppresses the effect of the fast free precession on the 
apsidal drift, see \S \S \ref{sect:timing}.

\begin{figure}
\plotone{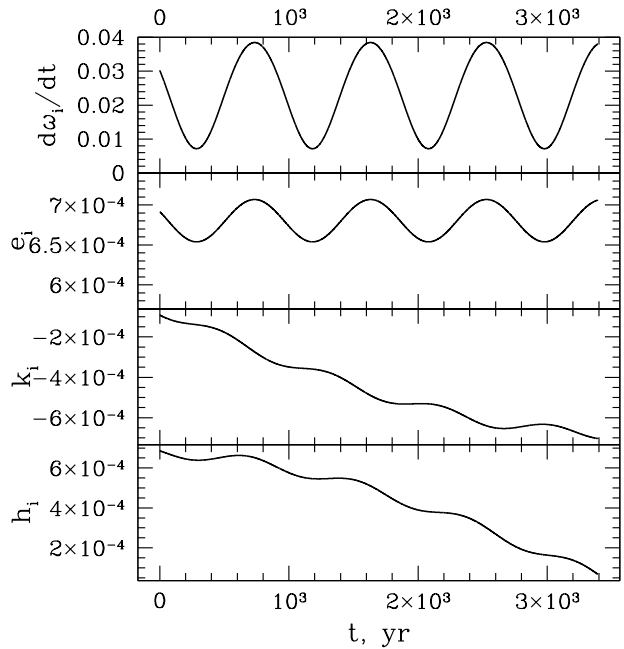}
\caption{Evolution of orbital elements for the inner pulsar on $\sim 10^3$ yr
timescales. Precession rate in the upper panel is in $^\circ$yr$^{-1}$.
\label{fig:inner_ev}}
\end{figure}


\section{Outer binary}  
\label{sect:outer}

\begin{figure}
\plotone{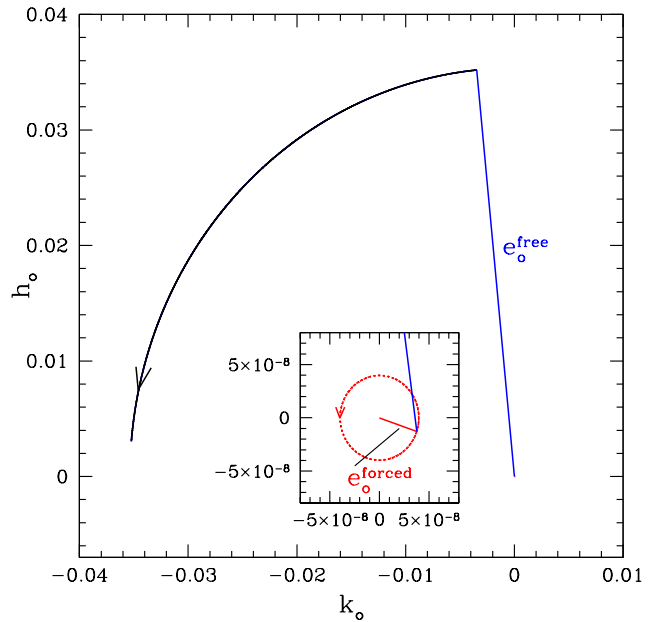}
\caption{Precession of the eccentricity vector of the outer binary
during the period indicated in Figure \ref{fig:outer_ev}. 
\label{fig:outer_prec}}
\end{figure}

For the outer binary the eccentricity vector is decomposed as
\ba
{\bf e}_o & = & {\bf e}_o^{\rm free}+{\bf e}_o^{\rm forced},
\label{eq:out_decomp}\\
{\bf e}_o^{\rm free} & = & e_{o,-}\left\{
\begin{array}{l}
\cos\left(g_- t+\beta_-\right) \\
\sin\left(g_- t+\beta_-\right)
\end{array}
\right\},
\label{eq:out_free}\\
{\bf e}_o^{\rm forced} & = & e_{o,+}\left\{
\begin{array}{l}
\cos\left(g_+ t+\beta_+\right) \\
\sin\left(g_+ t+\beta_+\right)
\end{array}
\right\}.
\label{eq:out_forced}
\ea
This decomposition is illustrated in Figure \ref{fig:outer_prec}.
Note that for the outer binary ${\bf e}_o^{\rm free}$ circulates at 
the rate $g_-$, while ${\bf e}_o^{\rm forced}$ rotates at 
the rate $g_+$.

\begin{figure}
\plotone{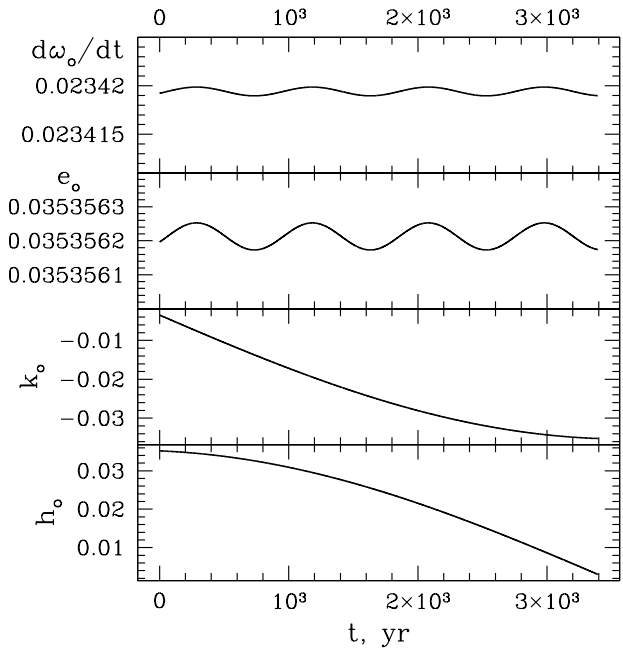}
\caption{Same as Figure \ref{fig:inner_ev} but for the outer binary.
\label{fig:outer_ev}}
\end{figure}

Compared to the inner binary, the relation between the free 
and forced eccentricity amplitudes is now reversed, 
and ${\bf e}_o^{\rm forced}$ is completely negligible
(by six orders of magnitudes) compared to ${\bf e}_o^{\rm free}$, 
see inset in Figure \ref{fig:outer_prec}. As a result, the evolution 
${\bf e}_o$ is accurately represented by the 
free precession at constant rate $g_-$. This is best seen in 
Figure \ref{fig:outer_ev}, where $|{\bf e}_o|$ varies 
at the level of $10^{-6}$, while $\dot\varpi_o$ varies at the
$10^{-5}$ level. This is also the reason why the ${\bf e}_o^{\rm free}$ 
(blue) and ${\bf e}_o$ (black) trajectories in Figure 
\ref{fig:outer_prec} fall on top of each other. 

The present day value of the precession rate for the outer WD is
\ba
\dot\varpi_o\approx 0.0234^\circ~\mbox{yr}^{-1},
\label{eq:dvarpi_o}
\ea
and is not very different from $\dot\varpi_i$.


\section{Discussion.}  
\label{sect:disc}

Small value of the forced eccentricity of the outer binary 
implies that its eccentricity vector ${\bf e}_o$ is always 
very well aligned with the forced eccentricity vector of 
the inner binary ${\bf e}_i^{\rm forced}$, see equations 
(\ref{eq:in_decomp})-(\ref{eq:in_forced}) and 
(\ref{eq:out_decomp})-(\ref{eq:out_forced}).
The only significant source of misalignment between ${\bf e}_i$ and 
${\bf e}_o$ is due to the non-zero value of 
${\bf e}_i^{\rm free}$. However, because 
$e_i^{\rm free}\ll e_i^{\rm forced}$ this
misalignment is also rather small. Thus, the eccentricity 
vectors of both binaries are guaranteed to be {\it locked in 
near-alignment}. 

As a result, the observed close apsidal alignment between the two 
binaries at present epoch ($|\varpi_i-\varpi_o|=1.9987^\circ$) does 
not require unique circumstances such as observing the system at 
a special moment of time. The inferred misalignment is in 
fact close to the maximum possible 
max$|\varpi_i-\varpi_o|=\mbox{asin}
\left|e_i^{\rm free}/e_i^{\rm forced}\right|
\approx 2.16^\circ$, given the smallness of 
$e_i^{\rm free}$. 

Collinearity of ${\bf e}_o$ and ${\bf e}_i^{\rm forced}$ also 
explains why $\dot\varpi_i$ and $\dot\varpi_o$ are of the same 
order of magnitude --- they would be exactly equal if 
${\bf e}_i^{\rm free}$ were zero. However, because 
${\bf e}_i^{\rm free}$ is not zero and period of free
precession of the inner binary is shorter that that of the 
${\bf e}_i^{\rm forced}$ circulation, we find that $\dot\varpi_i$
exhibits significant variability around the value corresponding 
to $\dot\varpi_o$, see Figure \ref{fig:inner_ev}. 

Note that these conclusions rely on the use of secular approximation
and strictly speaking apply to the orbital evolution of the system
on long ($\sim P_+$ and longer) intervals. In reality orbital 
parameters of both binaries will also oscillate on shorter timescales 
(e.g. $P_o$) due to the short-period terms in the expansion of 
the disturbing function (Murray \& Dermott 1999). We cannot 
capture such effects with our orbit-averaged approach. However, 
we expect the amplitude of short-term variations to be small 
compared to the secular ones. The general behavior 
of the system should still be reasonably well described by our 
results.


\subsection{Period-eccentricity relation}  
\label{sect:P-e}

Millisecond radio pulsars with WD companions are known to obey 
the so-called {\it eccentricity-period relation}, which states 
that the eccentricity of the neutron star (NS)-WD binary $e_b$ 
increases with its orbital period $P_b$ (Lorimer 2008). 
Phinney (1992) suggested 
that this correlation emerges at the last stage of the Roche 
lobe overflow (RLOF) leading to the formation of the WD: random 
density fluctuations in the envelope of the WD progenitor 
caused by the convective motions induce stochastic variations 
of the gravitational quadrupole tensor $Q_{ij}$ of the progenitor.
This results in random quadrupole accelerations acting on the 
NS and drives eccentricity of the binary. Random walk of $e_b$
is balanced by the tidal dissipation in the WD envelope, which 
results in a well-defined theoretical correlation between $e_b$ 
and $P_b$. This theory predicts, in particular, the rough 
equipartition between the kinetic energy of individual convective 
eddies in the envelope of the WD progenitor and the energy of 
eccentric motion of the binary. 

\begin{figure}
\plotone{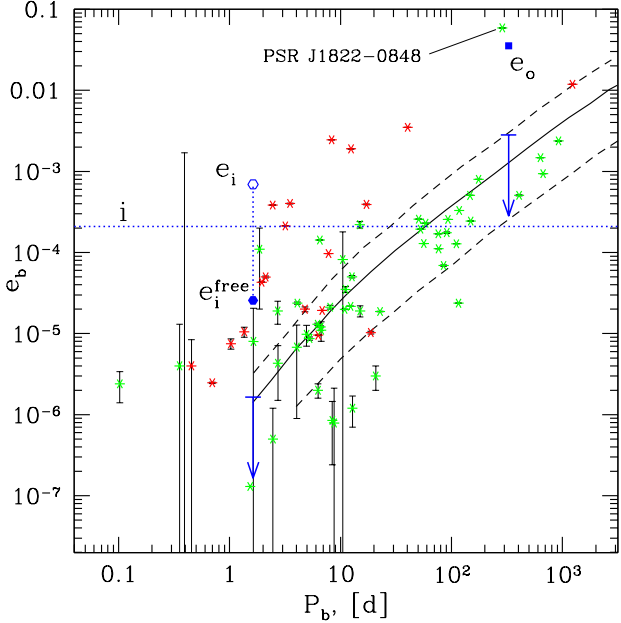}
\caption{
Eccentricity-period ($e_b-P_b$) relation for binary pulsars with 
WD companions. Red points are for systems with CO WDs, green are 
for systems with He WDs. Binaries comprising the system PSR 
J0337+1715 are indicated with blue points: inner eccentricity $e_i$ 
(open hexagon), inner free eccentricity $e_i^{\rm free}$ (filled 
hexagon), outer eccentricity $e_o$ (filled square), and mutual 
inclination $i$ (dotted line). Downward arrows are the theoretical 
upper limits on the mutual eccentricity given by equations 
(\ref{eq:Ie_b_i}) for the inner and (\ref{eq:Ie_b_o}) for the outer 
binaries. Solid curve shows theoretical 
$e_b-P_b$ relation from Phinney and Kulkarni (1994), with dashed 
curves encompassing $95\%$ deviations. 
\label{fig:ecc_per}}
\end{figure}

More recent compilations of the binary pulsar properties 
(Ng \etal 2014) show that many systems deviate from $e_b-P_b$ 
relation suggested by Phinney (1992) 
by orders of magnitude in $e_b$. Nevertheless, the general 
trend of $e_b$ increasing with growing $P_b$ is still 
observed, see Figure \ref{fig:ecc_per} in which we display 
properties of 72 binary pulsars with both CO and He WD 
companions. The data are from the ATNF pulsar 
catalogue\footnote{http://www.atnf.csiro.au/people/pulsar/psrcat/} 
(Manchester \etal 2005) and include only NS-WD systems which do not 
reside in globular clusters.

The system of PSR J0337+1715 provides two tests of the $e_b-P_b$
relation, for both binaries. This is because the system must have 
undergone {\it two} RLOF episodes, within the inner and outer 
binaries, to arrive to its present configuration with two WDs 
(TvdH14).

Present day eccentricity of the inner 
binary, $e_i\approx 6.9\times 10^{-4}$ is more than two orders 
of magnitude larger $e_b\sim 10^{-6}$ implied by the  $e_b-P_b$ 
relation of Phinney (1992), see Figure \ref{fig:ecc_per}. However, 
as mentioned in \S \ref{sect:inner}, almost all of this 
eccentricity is forced by the gravitational perturbations due 
to the outer WD. Only the free eccentricity $e_i^{\rm free}
=e_{i,+}\approx 2.6\times 10^{-5}$, which is much smaller than 
$e_i$, carries information about the initial conditions for the 
evolution of the inner binary. Even this low value is still 
about an order of magnitude higher than the prediction of 
Phinney (1992), see Figure \ref{fig:ecc_per}. However, same Figure
also shows several other NS-WD systems with $P_b$ of order several 
days and $e_b$ deviating from this relation upward by at least an 
order of magnitude. Thus, the inner binary of PSR J0337+1715, 
while disagreeing with the 
theoretical $e_b-P_b$ correlation, is not an extreme 
outlier in the general population of the short-period NS-WD binaries. 

For the outer binary the non-Keplerian gravitational perturbations 
due to the inner binary play negligible role and its free eccentricity
is essentially equal to its current eccentricity $e_0\approx 0.035$.
This is more than an order of magnitude higher than $e_b\sim 10^{-3}$ 
predicted by Phinney (1992) for the orbital period of $P_o=$327 d, 
see Figure \ref{fig:ecc_per}.
On the other hand, there is another pulsar (PSR J1822-0848, Lorimer
\etal 2006) in a similar orbit with $P_b=287$ d and high $e_b=0.059$, 
which also strongly deviates from the $e_b-P_b$ relation,
see Figure \ref{fig:ecc_per}. Thus, 
systems like the outer binary of PSR J0337+1715 
are not unique and have been previously known.


\subsection{Origin of the binary eccentricities}  
\label{sect:e_origin}

We now explore what do the measurements of 
binary eccentricities in the PSR J0337+1715 triple tell us about 
the past history of the system and the $e_b$-excitation mechanisms, 
focussing on the theory of Phinney (1992).

During the RLOF phase of the inner binary its eccentricity 
has already been driven by the outer WD. Tidal dissipation in 
the inner WD damps $e_i$ on the characteristic {\it 
circularization} timescale (Correia 2009; Correia \etal 2011)
\ba
t_{\rm tid}^e & = & \frac{2}{21}\frac{Q}{k_2}n_i^{-1}\frac{m_i}{m_p}
\left(\frac{a_i}{R_i}\right)^5
\label{eq:tidal}\\
& \approx & 1.6\times 10^5
\mbox{yr}\frac{Q/k_2}{10^7}\left(\frac{0.227}{R_i/a_i}\right)^5,
\nonumber
\ea
assuming present-day orbital parameters of the inner binary
and $R_i\approx 0.227a_i$ given by the Roche radius formula of 
Eggleton (1983) for $q=m_i/m_p=0.137$;  $k_2$ is the 
Love number and 
$Q$ is the tidal quality factor. This timescale is much 
longer than the characteristic secular timescale $P_+$ 
of the inner binary, see equation (\ref{eq:P_pm}). It has 
been shown by Mardling (2007) and Batygin \etal (2009)
that in these circumstances tidal dissipation drives
${\bf e}_i$  towards the {\it fixed point} state, in which 
$|{\bf e}_i|$ is finite and fixed and 
apsidal lines of the inner and outer binaries are strictly 
aligned. In this configuration the free eccentricity 
of the inner binary is damped to zero by tides. 

One can also show that as long as 
$\left(A_i-A_0\right) t_{\rm tid}^e\gg 1$ the forced 
eccentricity of the inner WD is given by the expression
computed in the absence of tidal dissipation. Moreover, the 
stochastic excitation of $e_i$ by the convective eddies has 
likely been negligible, since it did not result in large 
present day value of $e_i^{\rm free}$. Thus, during the RLOF 
phase $e_i$ should have been close to the present-day forced 
eccentricity $e_i^{\rm forced}=e_{i,-}\approx 6.8\times 10^{-4}$,
if the orbital parameters of the triple did not change since then.
The tidal dissipation rate due to this non-zero $e_i$ is too small to 
affect the semi-major axis of the inner binary --- $a_i$ would 
appreciably change only on timescale 
$\sim e_i^{-2}t_{\rm tid}^e\gtrsim 10^{11}$ yr. However, the 
persistent non-zero eccentricity at the RLOF stage might have 
interesting consequences for the dynamics of the mass transfer
through the inner Lagrange point.

As the inner WD loses its outer envelope and accretion onto the
NS ceases, $t_{\rm tid}^e$ becomes very long, about $2\times 10^{11}$ yr 
for the present-day inner WD radius $R_i\approx 0.091R_\odot$ 
(Ransom \etal 2014; Kaplan \etal 2014) and $Q/k_2=10^7$. Any
$e_i^{\rm free}$ that is imparted by e.g. the convective eddies 
in the vanishing envelope of the inner WD at the ``freeze-out'' 
moment, when $t_{\rm tid}^e$ becomes comparable to the internal 
evolution timescale of the inner WD, gets inherited by the binary 
until present time. Whatever process was responsible for the 
eccentricity excitation, it must have been very efficient at 
that moment since currently
measured $e_i^{\rm free}$ considerably exceeds the prediction 
of Phinney (1992).

Anomalously high eccentricity of the outer binary might be due to 
the same physical mechanism that excited $e_i$ but operating 
during the earlier RLOF phase of the outer WD (TvdH14). However, 
this is just a possibility. Since 
${\bf e}_o^{\rm forced}\ll {\bf e}_o^{\rm free}$, it is very hard 
to imagine that the inner binary has exchanged enough angular 
momentum with the outer one to excite $e_o$ to the 
present day value, especially in light of the evolutionary 
scenario favored by TvdH14, in which the inner binary has always been 
essentially circular after the birth of the outer WD. Thus, we 
conclude that non-zero $e_o$ was induced when the 
outer companion was turning into the WD, either via a very 
efficient version of the Phinney (1992) mechanism, or by some 
other means.


\subsection{Origin of the mutual inclination}  
\label{sect:inclin}

Triple nature of the PSR J0337+1715 system presents us with 
a unique chance to explore excitation of the epicyclic
motion in a NS-WD binary not only in the orbital plane but 
also {\it in the direction normal to it}. Measurement of the 
non-zero angle between the orbital planes of the two binaries, 
$i=1.2\times 10^{-2~\circ}\approx 2.1\times 10^{-4}$ (Ransom 
\etal 2014) has a potential to provide us with important clues 
to the processes sculpting the dynamical architecture of 
the system. 

This measurement is interesting because one expects $i$ to be 
reduced to zero during the RLOF phase of the outer binary, when
an accretion disk coplanar with it was engulfing the inner binary. 
During this stage that could have lasted for $\sim 17$ Myr 
(TvdH14) gravitational torques acting on the inner binary must
have reduced its inclination to zero: the angular momentum 
deposited into the disk by the inner binary, regardless of 
whether it was able to clear out an inner cavity in the disk or not, 
gets communicated by pressure and viscous stresses out to the 
outer binary, effectively damping their mutual inclination 
(Artymowicz 1994; Terquem 1998).

After the complete orbital alignment of the two binary planes, 
their non-zero mutual inclination could have been excited by the same 
underlying mechanism as the binary eccentricity, or a completely 
different
one. Here we will focus on the former possibility, i.e. 
epicyclic excitation by random forces due to stochastically 
generated quadrupole in the convective envelope of the WD 
progenitor (Phinney 1992). This mechanism should naturally drive 
out-of-plane motion. 

Indeed, the r.m.s. radial and
vertical components of the pulsar acceleration due to 
stochastically varying WD quadrupole are
\ba
a_r^{\rm rms}=\frac{3}{2}\frac{G Q_{rr}^{\rm rms}}{a^4},~~~
a_z^{\rm rms}=\frac{G Q_{rz}^{\rm rms}}{a^4},
\label{eq:accels}
\ea
where $Q_{rr}^{\rm rms}$ and $Q_{rz}^{\rm rms}$ are the r.m.s. 
values of the relevant diagonal and off-diagonal components of the 
WD quadrupole tensor. Since non-zero $Q_{ij}$ is due to a 
superposition of many randomly varying convective eddies, 
one can show (Phinney 1992) that 
$Q_{rz}^{\rm rms}=\left(3/4\right)^{1/2}Q_{rr}^{\rm rms}$,
so that 
\ba
\frac{a_z^{\rm rms}}{a_r^{\rm rms}}=3^{-1/2},
\label{eq:accel_rat}
\ea
i.e. inclination is driven somewhat less efficiently than 
eccentricity. 

Excitation of $e_b$ and mutual inclination $I$ by the WD 
quadrupole is balanced
by the tidal dissipation and one expects (Phinney 1992) 
their equilibrium values  to scale as 
$e_b\propto \left(t_{\rm tid}^e\right)^{1/2}a_r^{\rm rms}$, 
$I\propto \left(t_{\rm tid}^i\right)^{1/2}a_z^{\rm rms}$, where 
$t_{\rm tid}^e$ is given by equation (\ref{eq:tidal}) and 
$t_{\rm tid}^i$ is the inclination damping time. Thus,
one expects
\ba
\frac{I}{e_b}\approx \frac{a_z^{\rm rms}}{a_r^{\rm rms}}
\left(\frac{t_{\rm tid}^i}{t_{\rm tid}^e}\right)^{1/2}.
\label{eq:Ie_b}
\ea
Note that this relation does not make assumptions about
the physical nature of the random acceleration, however the 
value of $I/e_b$ depends on the knowledge of $t_{\rm tid}^i$.

Because of tidal dissipation, over the long time interval the 
hierarchical triple system like PSR J0337+1715 tends to 
converge to an equilibrium 
configuration in which the inner and outer orbits are coplanar 
and the inner WD obliquity $\theta_i$ --- the angle between 
its spin axis ${\bf S}_i$ and the orbital angular momentum 
${\bf L}_i$ of the inner binary --- is zero. In this state  
${\bf L}_i$ is aligned with the orbital angular momentum of 
the outer binary ${\bf L}_o$, which dominates over
${\bf L}_i$, and the WD rotation is synchronized with the
orbital motion (we neglect the eccentricity of the inner 
binary). 

Next we discuss the way in which convergence to 
this equilibrium state takes place and the value of 
inclination damping timescale $t_{\rm tid}^i$. We do this 
for two alternative scenarios of the inclination driving,
depending on whether it is excited by processes operating 
during the RLOF phase of the inner or outer binaries.


\subsubsection{Excitation of $I$ during the inner RLOF phase}  
\label{sect:inn_RLOF}

TvdH14 advocate an evolutionary 
scenario in which the inner WD undergoes RLOF 
{\it after} the outer one. We explore whether the mutual 
inclination could have 
been imprinted at the end of this phase, simultaneously with 
the free eccentricity excitation for the inner binary. 

Random torques driven by the convection in the WD progenitor 
excite both the mutual inclination of the two binaries 
$\delta I$ (assuming initial coplanarity) and the obliquity 
of the inner WD $\delta \theta$. Angular momentum conservation 
ensures that 
\ba
\delta\theta=\frac{|{\bf L}_i|}{|{\bf S}_i|}\delta I,
\label{eq:theta_I}
\ea
so that $\delta\theta\gg \delta I$ since 
$|{\bf L}_i|\gg|{\bf S}_i|$ for a synchronized WD progenitor. 
Obliquity of the WD damps on the {\it synchronization} 
timescale (Correia \etal 2011)
\ba
t_{{\rm sync},i} & = & 
\frac{2}{3}\frac{\xi Q}{k_2}n_i^{-1}\frac{m_i(m_p+m_i)}
{m_p^2}\left(\frac{a_i}{R_i}\right)^3
\label{eq:t_theta}\\
& \approx & 600~\mbox{yr}~
\frac{Q/k_2}{10^7}\frac{\xi}{0.01}\frac{0.227}{R_i/a_i},
\nonumber
\ea
which is much shorter than the circularization timescale 
$t_{\rm tid}^e$, see equation (\ref{eq:tidal}). Because
of the relation (\ref{eq:theta_I}) mutual inclination 
decays due to tides on the same short timescale 
$t_{{\rm sync},i}$, simultaneous with the excitation by
random quadrupole torques. Substituting $t_{{\rm sync},i}$ for 
$t_{\rm tid}^i$ in equation (\ref{eq:Ie_b}) we find
\ba
\frac{I}{e_i} & \approx & \chi
\left(7\xi\frac{m_i+m_p}{m_p}\right)^{1/2}\frac{R_i}{a_i}=
\chi\left(7\frac{|{\bf S}_i|}{|{\bf L}_i|}\right)^{1/2}
\label{eq:Ie_b_i}\\
& \approx & 0.064~\chi\left(\frac{\xi}{0.01}\right)^{1/2}
\frac{R_i/a_i}{0.227},
\nonumber
\ea
where we defined $\chi\equiv a_z^{\rm rms}/a_r^{\rm rms}$. 

Equation (\ref{eq:Ie_b_i}) implies that even for $\chi=1$ 
the mutual inclination $I$ excited during the inner binary RLOF 
should be {\it much smaller} than the free eccentricity of 
the inner WD, at the level of $\lesssim 10^{-6}$. This is because
spin angular momentum of the pulsar companion is much smaller
than the orbital angular momentum of the binary.
However, the present day value of $I=i$ is about an order of 
magnitude {\it higher} than $e_i^{\rm free}$, 
$i/e_{i,+}\approx 8.2$, exceeding the prediction 
(\ref{eq:Ie_b_i}) by more than two orders of magnitude, see 
the upper limit at $P_i$ in Figure \ref{fig:ecc_per}. 

The discrepancy can in fact be even worse as the assumption 
of $\chi\sim 1$ suggested by the estimate (\ref{eq:accel_rat}) 
may be too optimistic. Indeed, the final value of $i$ 
gets established at the freeze-out time when 
$t_{{\rm sync},i}$ becomes comparable to the WD evolution 
timescale at a stage when mass transfer stops and $R_i$ becomes
smaller than the Roche radius. Since $t_{{\rm sync},i}\ll t_{\rm tid}^e$ 
and $t_{{\rm sync},i}$ scales with $R_i/a_i$ slower than 
$t_{\rm tid}^e$, the inclination freeze-out must occur 
{\it considerably later} than the eccentricity freeze-out 
of the inner binary. At that stage convective envelope of the
WD progenitor is less massive than at $e_i$ freeze-out, resulting 
in less vigorous quadrupole fluctuations and $a_z^{\rm rms}$ 
being likely smaller than $a_r^{\rm rms}$ was when $e_i$ attained 
its final value. 

As a result, one should expect  $\chi\ll 1$,
exacerbating the discrepancy between the theoretical and 
measured values of $i$. For these reasons we believe that
the present day mutual inclination could not have been 
established at the end of the RLOF phase of the inner binary.

In conclusion we would like to address a subtle point related
to the energy equipartition between individual convective 
eddies and the binary epicyclic motion. Phinney (1992) has
shown that the mean energy of the random epicyclic motion 
in the binary plane $E_e=(1/2)\mu v_K^2 e_b^2$ ($\mu$ is the 
reduced mass, $v_K$ is the Keplerian speed) is of order the 
kinetic energy of an individual eddy driving $Q_{ij}$ 
fluctuations. Equation (\ref{eq:Ie_b_i}) then suggests that 
the mean energy of the random epicyclic motion normal to 
the binary plane $E_i=(1/2)\mu v_K^2 I^2\ll E_e$ and is not
in equipartition with individual convective eddies. 

This paradox is easily resolved when one notices that 
inclination variations unavoidably cause much larger 
obliquity variations of the WD progenitor, see equation 
(\ref{eq:theta_I}). Because both are driven by equal (but
opposite) torques, the energy stored in the obliquity 
wobble of the pulsar companion $E_\theta$ must be larger than 
$E_i$ by a factor $\delta \theta/\delta I=|{\bf L}_i|/|{\bf S}_i|\gg 1$.
Using equations (\ref{eq:accel_rat}) and (\ref{eq:Ie_b_i})
one then trivially finds that $E_\theta\approx (7/3)E_e$.
Thus, the full binary energy $E_\theta+E_i\approx E_\theta$ 
associated with the out-of-plane torques is in fact in 
equipartition with the kinetic energy of individual 
eddies in the WD progenitor envelope.


\subsubsection{Excitation of $I$ during the outer RLOF phase}  
\label{sect:out_RLOF}

Results of \S \ref{sect:inn_RLOF} can be trivially extended to
study the possibility of the mutual inclination excitation by
the random $Q_{ij}$ fluctuations during the RLOF phase of 
the {\it outer} binary. The same logic applies in that case as well
if we consider inner binary as a point mass (which is reasonable 
for an hierarchical system) and replace $m_p\to m_p+m_i$ and 
$m_i\to m_o$ in all equations. As a result, we find that random
fluctuations of the mutual inclination during the outer RLOF 
get damped on timescale
\ba
t_{{\rm sync},o} & = & 
\frac{2}{3}\frac{\xi Q}{k_2}n_o^{-1}\frac{m_om_3}
{(m_p+m_i)^2}\left(\frac{a_o}{R_o}\right)^3
\label{eq:t_theta_o}\\
& \approx & 1.6\times 10^5~\mbox{yr}
\frac{Q/k_2}{10^7}\frac{\xi}{0.01},
\nonumber
\ea
assuming the present day masses of all components. If we follow 
TvdH14 and adopt $m_p=1.3M_\odot$ and 
$m_i=1.12M_\odot$ (mass of the inner WD progenitor) during the 
outer binary RLOF, prior to the inner RLOF episode, this 
timescale does not change significantly, 
$t_{{\rm sync},o}\approx 10^5$ yr. This is shorter than the expected
duration of the outer RLOF phase, $\sim 17$ Myr (TvdH14). 

Analogously, equation (\ref{eq:Ie_b_i}) becomes
\ba
\frac{I}{e_o} & \approx & \chi
\left(7\xi\frac{m_3}{m_i+m_p}\right)^{1/2}\frac{R_o}{a_o}
\label{eq:Ie_b_o}\\
& \approx & 0.08~\chi\left(\frac{\xi}{0.01}\right)^{1/2}
\frac{R_o/a_o}{0.268},
\nonumber
\ea
where we used $R_o=0.268a_o$ for $q=m_o/(m_p+m_i)=0.251$. 
Adopting the parameters of the inner binary before its RLOF 
phase we get $q=m_o/(m_p+m_i)=0.17$, $R_o=0.24a_o$,
and essentially the same value of $I/e_o$.

This estimate shows that if the observed outer binary eccentricity
$e_o\approx 0.035$ was excited by randomly varying $Q_{ij}$ of
the WD progenitor at the end of the RLOF phase, then the same 
process must have given rise to the mutual inclination 
$I\approx 2.8\times 10^{-3}\chi\left(\xi/0.01\right)^{1/2}$. 
This can be easily reconciled with the present day value of 
$i=2.1\times 10^{-4}$ (see the upper limit on $I$ at $P_o$ 
in Figure \ref{fig:ecc_per}) 
if either $\chi\lesssim 10^{-2}$, which 
is quite plausible for an extended WD progenitor in long-period 
orbit, or $\chi\lesssim 1$, which is also expected, as we described 
in \S \ref{sect:inn_RLOF}.

However, this comparison of $I$ and the current mutual inclination 
$i$ is meaningful only if $I$ did not change since the outer binary
RLOF phase, which may have occurred $\sim 5$ Gyr ago (TvdH14). 
Mutual inclination would decay because for non-zero $I$ the steady
state obliquity $\theta_i$ of the inner pulsar companion 
(WD or its progenitor) does not converge to $\theta_i=0$.
Instead, tidal dissipation in the companion drives the system 
towards the so-called Cassini state (Colombo 1966; Peale 1969), 
in which the companion spin vector ${\bf S}_i$, the angular 
momentum of the inner binary ${\bf L}_i$, and the total angular 
momentum of the system ${\bf J}$ are coplanar and precess at the 
same rate $A_i^{\rm sec}$. Convergence of the companion spin to 
this configuration happens on 
relatively short synchronization timescale $t_{{\rm sync},i}$ 
given by equation (\ref{eq:t_theta}).

Obliquity in the low-obliquity Cassini state (the one relevant 
for our purposes) is non-zero and is given in the case of weak 
dissipation by $\theta_i^{\rm Cas}\approx \lambda_i^{-1}I$ 
(Ward \& Hamilton 2004), where the dimensionless parameter 
\ba
\lambda_i & = & \frac{2}{3}\frac{k_2}{\xi}
\frac{m_p(m_p+m_i)}{m_i m_o}\left(\frac{R_i}{a_i}\right)^3
\left(\frac{a_o}{a_i}\right)^3
\label{eq:lambda_i}\\
& \approx & 1.1\times 10^4\frac{k_2}{\xi}
\left(\frac{R_i/a_i}{0.227}\right)^3
\nonumber
\ea
is the ratio of the pulsar companion spin precession rate to 
the nodal precession rate of the inner binary. Tidal dissipation 
affects the value of $\theta_i^{\rm Cas}$ (Fabrycky \etal 2007) 
but for $A_i^{\rm sec}t_{\rm tid}^e \gg 1$ the effect is small 
and will be neglected. The numerical estimate in equation 
(\ref{eq:lambda_i}) applies during the inner binary RLOF, when 
$\lambda_i\gg 1$. During other evolutionary phases $R_i/a_i$
is smaller and $\lambda_i$ is lower. But in any case it is 
reasonable to expect $\lambda_i\gtrsim 1$ so that 
$\theta_i^{\rm Cas}\lesssim I$. 

Non-zero obliquity in presence of tidal dissipation results in 
a torque acting on the inner binary, which over time damps 
the mutual inclination $I$ of the two orbits. Its evolution 
is described by (Correia \etal 2011) 
\ba
\frac{dI}{dt}= - \frac{\sin\theta_i}{t_I},
\label{eq:Iev}
\ea
where $t_I\equiv 7t_{\rm tid}^e\gg t_{{\rm sync},i}$. Thus, 
${\bf S}_i$ settles into a Cassini state before $I$ has a 
chance to change. Plugging the dissipationless 
$\theta_i^{\rm Cas}\approx \lambda^{-1}I$ for the Cassini state 
into equation (\ref{eq:Iev}) we find that subsequently $I$
decays on a timescale 
\ba
t_I^{\rm Cas} & = & \lambda_i t_I=\frac{4}{9}\frac{Q}{\xi}n_i^{-1}
\frac{m_i+m_p}{m_o}\left(\frac{a_o}{a_i}\right)^3
\left(\frac{a_i}{R_i}\right)^2,
\label{eq:t_I}\\
& \approx & 12~\mbox{Gyr}~\frac{Q/\xi}{10^7}
\left(\frac{0.227}{R_i/a_i}\right)^2,
\nonumber
\ea
see equation (\ref{eq:tidal}). 

Now we need to distinguish two possibilities. First, TvdH14 
favor evolutionary scenario in which the inner binary underwent 
RLOF phase {\it after} the outer one. Given the scaling 
$t_I^{\rm Cas}\propto R_i^{-2}$ one expects fastest $I$ decay
to occur during the RLOF. But even then the estimate (\ref{eq:t_I})
shows that $t_I^{\rm Cas}$ is much longer than the expected duration
of the inner RLOF episode, $\sim 2$ Gyr (TvdH14). As a result, mutual
inclination does not decay during this phase by more than $\sim 20\%$.

Before and after that phase, the inner WD or its progenitor have 
$R_i/a_i$ smaller than during the overflow, resulting in much longer
$t_I^{\rm Cas}$. Thus, $I$ stays constant during these 
time intervals.

Second, TvdH14 also do not exclude the possibility of the outer WD 
to be the {\it last} one to form via the RLOF, after the inner 
WD has already formed. In this case $R_i/a_i$ is always very small 
and $t_I^{\rm Cas}$ is much longer than the lifetime of the system. 

High degree of the orbital coplanarity of the system also
strongly argues against the outer binary eccentricity $e_o$ being 
excited by some external process, e.g. gravitational perturbation 
by a passing star. Such perturbation is expected to excite $I$ 
on par with $e_o$ so that an encounter driving $e_o$ two orders 
of magnitude stronger than $i$ seems extremely improbable.
Needless to say, stellar encounter is in any case highly 
unlikely for an object in the field as excitation of 
$e_o=0.035$ would require stellar passage within 
$\sim 10$ AU from the system (Heggie \& Rasio 1996).   

Thus, we conclude that the combination of arguments 
presented here and in \S \ref{sect:inn_RLOF} does not contradict  
the scenario in which the present day mutual inclination 
$i$ was excited by convectively-driven random $Q_{ij}$ 
fluctuations at the end of the outer binary RLOF phase,
simultaneous with the excitation of its eccentricity $e_o$.
On the other hand, we cannot exclude the possibility that $I$ and/or 
$e_o$ were produced by a mechanism completely unrelated
the stochastic $Q_{ij}$ variations of the WD progenitor, 
e.g. as a result of incomplete damping of the mutual misalignment by disk 
torques during the outer binary RLOF. This possibility may 
be hinted at by the large difference between $e_o$ and the 
prediction of Phinney (1992), see \S \ref{sect:P-e}. 
Finally, the weak decay of mutual inclination after its excitation 
in the outer binary does not allow us to distinguish between
the two alternative evolutionary scenarios presented in TvdH14,
in which the outer WD is either the first or last one to 
form.


\subsection{Implications for timing measurement}  
\label{sect:timing}

Strong Newtonian three-body coupling between the components of 
PSR J0337+1715 system should make the measurement of 
the general relativistic effects in it difficult, even despite the high 
timing accuracy of this pulsar (Ransom \etal 2014). In 
particular, as shown in \S \ref{sect:inner} apsidal precession 
of the inner binary is no longer set by the GR precession 
alone but is instead strongly suppressed by the three-body 
effects, so that $\dot\varpi_i\approx 0.24 \dot\varpi_{\rm GR}$.

At the same time, we find that measurement of $\dot\varpi_i$ is
quite sensitive to the actual value of $\dot\varpi_{\rm GR}$.
Artificially varying $\dot\varpi_{\rm GR}$ by $1\%$ results 
in roughly $4\%$ variation of $\dot\varpi_i$. This is especially 
surprising given that $\dot\varpi_{\rm GR}$ provides only a 
$29\%$-contribution to $A_i$, see equations 
(\ref{eq:A_in})-(\ref{eq:omGR}). Such a disproportionate 
response can be understood if one manipulates the expression for
$\dot\varpi_i$ using solutions (\ref{eq:h_i})-(\ref{eq:k_i}) 
and the fact that $e_{i,+}/e_{o,+}\gg e_{i,-}/e_{o,-}$,
$g_+\approx A_i$, $g_-\approx A_o$. As a result, one can 
write the following expression for the apsidal rate of the 
inner binary at the present time:
\ba
\dot\varpi_i(0)\approx A_i+B_i\frac{{\bf e}_i\cdot {\bf e}_o}
{e_i^2},
\label{eq:varpinow}
\ea
where ${\bf e}_i$ and ${\bf e}_o$ are the eccentricity vectors 
of the inner and outer binaries at present day.
 
This formula clearly shows the near cancellation of the two 
large contributions (two terms in the r.h.s.), since 
$\dot\varpi_i(0)\ll A_i$ according to our estimate 
(\ref{eq:dvarpi_i}). Because of that even a small variation 
in the value of $A_i$ in equation (\ref{eq:varpinow}), e.g. 
due to the deviation of $\dot\varpi_{\rm GR}$ from the
GR prediction (Damour \& Taylor 1992), drives large 
change in the value of $\dot\varpi_i$. Thus, given a 
significantly accurate measurement of $\dot\varpi_i$ 
one may still be able to use timing of PSR J0337+1715
to constrain non-GR contributions to the apsidal precession
rate of the inner binary, even in presence of the three-body 
effects. The problem may be in measuring $\dot\varpi_i$ 
accurately enough, given the low value of $e_i$ (Ransom, 
private communication).

Another complication for timing may arise from the short-term 
effects, neglected in our calculations. Small variations of the
orbital parameters on the orbital timescales of both binaries 
may be at least partly degenerate with the 
post-Newtonian contributions to timing (Damour \& Deruelle 1986;
Damour \& Taylor 1992). This is likely to make the problem of 
extracting post-Newtonian parameters of the system quite 
challenging.


\section{Summary.}  
\label{sect:summ}

We studied secular dynamics of a recently discovered 
triple system containing radio pulsar PSR J0337+1715 and two WDs. 
Three body interactions play important role in the dynamics of 
this hierarchical system. Gravitational perturbations due to the 
outer WD endow the inner binary with a large forced eccentricity 
and strongly suppress its apsidal precession rate $\dot\varpi_i$, 
to about $24\%$ of the rate predicted by 
the general relativity. At the same time,  $\dot\varpi_i$ is 
still quite sensitive to the non-Newtonian contribution to the 
precession rate and may be used to constrain alternative gravity 
theories provided that a good measurement of $\dot\varpi_i$ is 
available. Small value of the free eccentricity 
$e_{i}^{\rm free}\approx 2.6\times 10^{-5}$ of the inner binary 
compared to the forced one and vanishing forced eccentricity of the 
outer, eccentric binary naturally result in apsidal near-alignment 
of the two binaries. 

These results help understand the evolutionary pathways 
leading to the current dynamical configuration of the system.
Free eccentricity of the inner binary exceeds the value predicted 
by the theoretical period-eccentricity relation of Phinney (1992) 
by more than an order of magnitude. Thus, both the inner and outer 
binaries of this hierarchical triple belong to the population of 
the NS-WD binaries exhibiting eccentricity in excess of that 
predicted by Phinney (1992). 

Moreover, the non-zero mutual 
inclination $i$ between the two orbits provides us with a unique 
chance to explore the {\it anisotropy} of epicyclic motion 
excitation in NS-WD binaries. We find it unlikely
that $i$ was excited, simultaneous with eccentricity, by random 
torques due to convective eddies in the WD progenitor during the 
RLOF in the inner binary. 

On the other hand, the smallness of the mutual inclination compared 
to the outer binary eccentricity argues in favor of simultaneous
excitation of $i$ and $e_o$ by random torques during the outer 
binary RLOF. Inclination driven during this phase should persist 
until now despite the complexity of the intermediate evolutionary 
stages. These facts may provide useful clues for understanding 
epicyclic excitation in the NS-WD binaries in general.


\acknowledgements

I am grateful to Scott Ransom, Jihad Touma, and Jeremy 
Goodman for useful discussions and to Cherry Ng for providing 
the data on theoretical $e_b-P_b$ relation.


\end{document}